\documentclass[%
 reprint,%
 amssymb, amsmath,
 aps,cha,prc%
superscriptaddress
]{revtex4-2}
\usepackage[colorlinks=true,linkcolor=blue]{hyperref}%
\usepackage{physics}
\usepackage{graphicx}
\usepackage[T1]{fontenc}
\usepackage{ulem}
\usepackage{color}

\begin{document}

\title{Implications of parity-violating electron 
	scattering experiments on $^{48}$Ca (CREX) and $^{208}$Pb (PREX-II) for nuclear energy density functionals}%

\author{Esra Y\"uksel}
\email{e.yuksel@surrey.ac.uk}
\affiliation{Department of Physics, University of Surrey, Guildford, 
             Surrey, GU2 7XH, United Kingdom}

\author{Nils Paar}
\email{npaar@phy.hr}
\affiliation{Department of Physics, Faculty of Science, University of Zagreb, Bijenicka cesta 32, 10000 Zagreb, Croatia}

\begin{abstract}
Recent precise parity-violating electron scattering experiments 
on $^{48}$Ca (CREX) and $^{208}$Pb (PREX-II) provide a new insight on the formation of neutron skin in nuclei.
Within the energy density functional (EDF) framework, we investigate the implications of CREX and PREX-II data on nuclear matter symmetry energy and isovector properties of finite nuclei: neutron skin thickness and dipole polarizability. The weak-charge form factors from the CREX and PREX-II experiments are employed directly in constraining the relativistic density-dependent point 
coupling EDFs. The EDF established with the CREX data acquires considerably smaller values
of the symmetry energy parameters, neutron skin thickness and dipole polarizability both for $^{48}$Ca and $^{208}$Pb, in comparison to the EDF obtained using the PREX-II data, and previously established EDFs.
Presented analysis shows that CREX and PREX-II experiments could not provide consistent constraints for the isovector sector of the EDFs, and further theoretical and experimental studies are required.
\end{abstract}

\maketitle

The nuclear equation of state (EOS) is essential 
for understanding the properties of strongly interacting many-body systems like atomic nuclei and neutron stars \cite{doi:10.1146/annurev-nucl-102711-095018,RevModPhys.89.015007,ROCAMAZA201896}. Constraining the density dependence of the nuclear symmetry energy, which is a key feature of the nuclear EOS, represents a long-standing and unresolved question in nuclear physics and astrophysics \cite{ROCAMAZA201896}. The fundamental source of this challenge is that the nuclear symmetry energy 
cannot be determined directly by experiment, thus it is necessary to identify and use relevant observables on finite nuclei to constrain their values \cite{ROCAMAZA201896}.
The neutron skin thickness ($\Delta R_{np}$) \cite{PhysRevLett.85.5296,PhysRevC.86.015803,Tamii_2014,PhysRevC.81.051303, PhysRevC.64.027302}, dipole polarizability ($\alpha_{D}$) \cite{PhysRevC.81.051303,PhysRevC.85.041302,PhysRevC.88.024316,PhysRevC.92.064304}, and neutron star mass-radius \cite{Lattimer_2014,BURGIO2021103879} have been established as key observables to constrain the isovector channel of the nuclear energy density functional (EDF) and the symmetry energy parameters of the nuclear EOS around the saturation density.

The recent precise parity-violating electron scattering experiments 
on $^{48}$Ca (CREX) \cite{PhysRevLett.129.042501} and $^{208}$Pb (PREX-II) \cite{PhysRevLett.126.172502} provide new insight into
the neutron skin thickness in nuclei. Through the measurement of the
parity violating asymmetry $A_{PV}$, these experiments allow to determine
the nuclear weak-charge form factor $F_W$ that is also strongly correlated with the density dependence of the symmetry energy and the neutron skin thickness of nuclei, hence providing an important quantity to probe the isovector channel of the EDFs \cite{PhysRevC.88.034325}. The weak charge form factor $F_W$ is obtained with the Fourier transform of the weak charge density for a given momentum transfer. Since the charge density is known experimentally, the Coulomb distortions can also be corrected accurately. \cite{PhysRevC.63.025501}.
Therefore, the parity-violating electron scattering experiments provide precise and model-independent data with small uncertainties for the nuclear weak-charge form factor $F_W$, that could be used in constraining the EDFs \cite{PhysRevC.88.034325,PhysRevC.63.025501}.
The CREX \cite{PhysRevLett.129.042501} and PREX-II \cite{PhysRevLett.126.172502} experiments reported the weak-charge form factors of $^{48}$Ca and $^{208}$Pb as $F_{W}$(q=0.8733 $fm^{-1}$)= 0.1304 $\pm$  0.0052(stat)$\pm$0.0020(syst) and $F_{W}$(q=0.3978 $fm^{-1}$)= 0.368$\pm$ 0.013 (exp.), respectively. The measured parity 
violating asymmetry $A_{PV}$ in $^{48}$Ca and $^{208}$Pb has recently been
analyzed using EDFs, indicating that there are difficulties to 
describe $A_{PV}$ simultaneously in both nuclei \cite{PhysRevLett.127.232501,https://doi.org/10.48550/arxiv.2206.03134}.

In this letter, we establish effective interactions based on the relativistic EDF with density-dependent point couplings, using recent nuclear weak-charge form factor $F_W$ data from CREX \cite{PhysRevLett.129.042501} and PREX-II \cite{PhysRevLett.126.172502} experiments alongside with the selected ground state properties of nuclei. 
In this way, we aim to reveal the implications of the
CREX and PREX-II experiments on the properties of finite nuclei and nuclear matter,
especially the symmetry energy and its slope around the saturation density.
By calculating the relevant nuclear observables with the new interactions, in particular the neutron skin thickness and dipole polarizability, we discuss their relationship with the recent experimental data from parity violating electron scattering.

\begin{table*}[ht!]
\centering
\renewcommand\arraystretch{1.5}
\caption{The nuclear matter properties at saturation density for the DDPC-CREX, DDPC-PREX, and DDPC-REX interactions. The properties for the DD-PC1 \cite{PhysRevC.78.034318} and DD-PCX \cite{PhysRevC.99.034318} interactions are also given for comparison. The uncertainties of the obtained values are provided within the parenthesis.}
\begin{tabular}[t]{lccccc}
\hline
&E/A (MeV) & $m_{D}^{*}/m$ &$K_{0}$ (MeV) &$J$ (MeV) & $L$ (MeV) \\
\hline
DDPC-CREX & -15.989(16) & 0.5672(5)  & 225.48(1.55) & 27.01(23) & 19.60(1.01)  \\
DDPC-PREX & -16.108(19) & 0.5680(7)  & 235.41(2.42) & 36.18(0.80) & 101.78(9.34)  \\
DDPC-REX & -16.019(16) &0.5696(5)  & 242.95(76) & 28.86(0.33) & 30.03(2.06)    \\
DD-PC1 & -16.061 & 0.580  &  230.0 & 33.0 & 70.1       \\
DD-PCX & -16.026(18) & 0.5598(8) & 213.03(3.54) & 31.12(32) & 46.32(1.68) \\  
\hline
\label{nmp}
\end{tabular}
\end{table*}%

Theoretical framework employed in this study is based on the relativistic EDF, 
where the self-consistent solution of relativistic single-nucleon Kohn-Sham 
equations provides the nuclear ground state density and energy \cite{PhysRev.140.A1133, RevModPhys.71.1253}. In the formulation given by the Lagrangian density, we use an effective interaction between nucleons described with four fermion contact interaction terms, including isoscalar-scalar, isoscalar-vector, isovector-vector and isospace-space channels \cite{PhysRevC.78.034318,NIKSIC20141808}. It includes free nucleon terms, point coupling interaction terms, coupling of protons to the electromagnetic field, and the derivative term accounting for the leading effects of finite-range interactions. For a quantitative description of nuclear density distribution and radii, the derivative terms are also necessary \cite{PhysRevC.78.034318}. The couplings in the
interaction terms $\alpha_i (\rho)$ include explicit density dependence \cite{NIKSIC20141808}. The point coupling model includes 10 parameters ($a_{S}$, $b_{S}$, $c_{S}$, $d_{S}$, $a_{V}$, $b_{V}$, $d_{V}$, $b_{TV}$, $d_{TV}$ and $\delta_S$) \cite{NIKSIC20141808}. The relativistic Hartree-Bogoliubov (RHB) model \cite{NIKSIC20141808,VRETENAR2005101,PhysRevC.78.034318} is used to describe open-shell nuclei, including the pairing field formulated using separable pairing force, which also contains two parameters for the proton and neutron pairing strengths ($G_{p}$ and $G_{n}$) \cite{PhysRevC.80.024313}.

In this work, we employed the RHB model to constrain 12 model parameters by minimizing the $\chi^2$ objective function with a set of observables on nuclear properties \cite{NIKSIC20141808, Roca_Maza_2015, PhysRevC.99.034318}. In order to benchmark the role of CREX and PREX-II data, in the $\chi^2$ minimization we used the same nuclear ground state properties as in recent optimization of the DD-PCX interaction, i.e., the binding energies (34 nuclei), charge radii (26 nuclei), and mean pairing gaps (15 nuclei) (see supplementary material \cite{SM} for the details about selected nuclei and their properties). In the optimization of the new functionals we also used the latest nuclear weak-charge form factors $F_W$ obtained from the CREX ($^{48}$Ca) \cite{PhysRevLett.129.042501} and PREX-II ($^{208}$Pb) \cite{PhysRevLett.126.172502} experiments. 
First we have established two functionals, DDPC-CREX and DDPC-PREX, optimized by using nuclear properties given above and the weak-charge form factor data of $^{48}$Ca and $^{208}$Pb, respectively. The adopted errors for $F_W$ are taken as 0.5\%, that is relatively tight in order to reproduce well the experimental $F_W$ values. Next, an additional functional DDPC-REX is established, by employing weak-charge form factor data of both $^{48}$Ca and $^{208}$Pb, with the errors taken as 2.0\%, to provide more flexibility to accommodate $F_W$ data for the two nuclei. Following the optimization of the interactions, the statistical uncertainties of the model parameters are estimated using the co-variance analysis \cite{Dobaczewski2014}. The parameters of these three new DDPC functionals and their uncertainties are given in Supplemental material \cite{SM}.

In Table \ref{nmp}, we present the nuclear matter properties: energy per nucleon $E/A$, effective mass ($m_{D}^{*}/m$), incompressibility $K_{0}$, symmetry energy $J$ and slope of the symmetry energy $L$ at saturation density \cite{ROCAMAZA201896}, for new DD-PC functionals in comparison to the properties of the previously established DD-PC1 \cite{PhysRevC.78.034318} and DD-PCX \cite{PhysRevC.99.034318} interactions. We note that the symmetry energy parameters $J$ and $L$ of the DDPC-CREX functional are considerably lower than those of the DDPC-PREX, while the DDPC-REX acquires intermediate values between the two former functionals. While the $J$ and $L$ values for the DDPC-CREX functional are lower than those of DD-PC1 and DD-PCX, the DDPC-PREX functional predicts much higher values. Our DDPC-PREX results are also consistent with the findings in Ref. \cite{PhysRevLett.126.172503}, where $J=$38.1$\pm$4.7 MeV and $L=$106$\pm$37 MeV are obtained using the PREX-II data. The symmetry energy
parameters for DDPC-CREX and DDPC-PREX interactions are outside
rather broad ranges of their values obtained in the EDF analysis of
dipole polarizability, $J=30-35$ MeV and $L=20-66$ MeV \cite{PhysRevC.92.064304}, and also at variance with the findings of previous studies (see Ref. \cite{ROCAMAZA201896} and the references therein.)
Therefore, already at the level only of the nuclear matter properties, we observe considerable inconsistencies between the symmetry energy properties of new functionals constrained by weak-charge form factor data and previously established functionals that are known as successful in the description of a variety of nuclear properties, including DD-PCX \cite{PhysRevC.99.034318} with carefully adjusted isovector channel using nuclear observables such as dipole polarizability.

\begin{figure} [ht!]
	\centering
	\includegraphics[width=\linewidth]{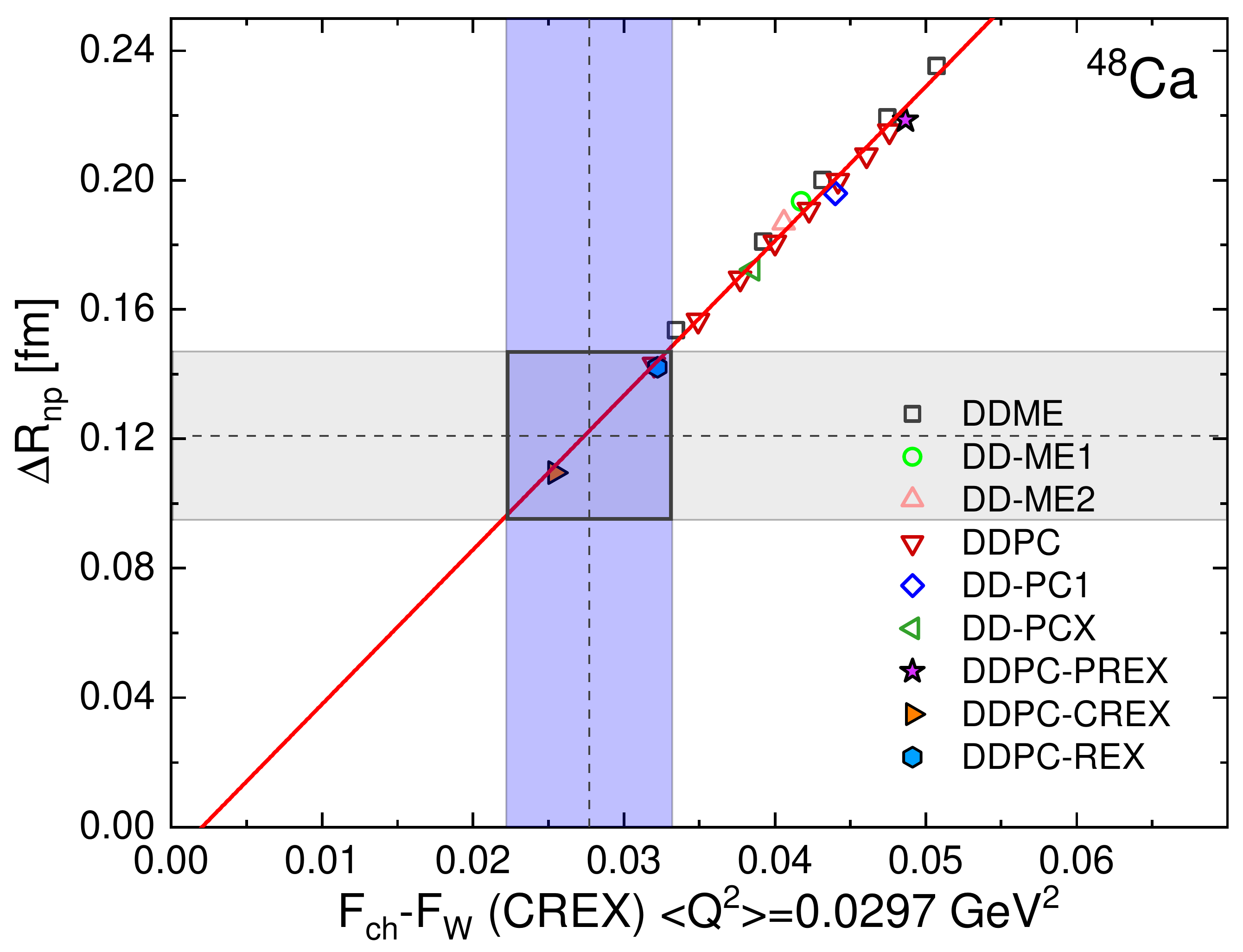}
	\includegraphics[width=\linewidth]{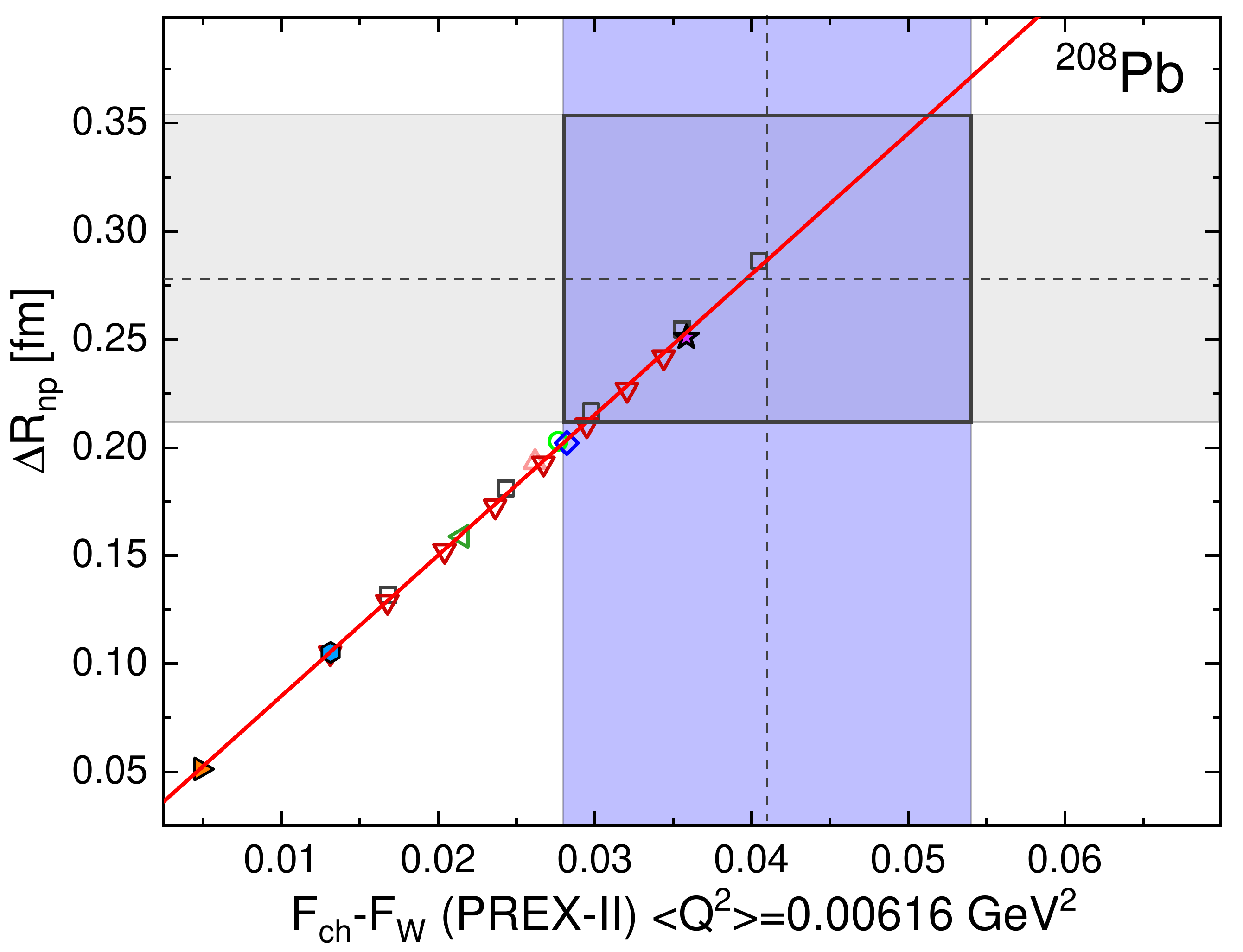}
	\caption{The neutron skin thickness $\Delta R_{np}$ of $^{48}$Ca (upper panel) and $^{208}$Pb  (lower panel) as a function of the form factor difference $F_{ch}-F_W$ using relativistic energy density functionals. The experimental data from PREX-II  \cite{PhysRevLett.126.172502} and CREX \cite{PhysRevLett.129.042501} are denoted with vertical and horizontal bands.}\label{fig:1}
\end{figure}

\begin{figure*} [ht!]
	\centering
	\includegraphics[width=0.8\linewidth]{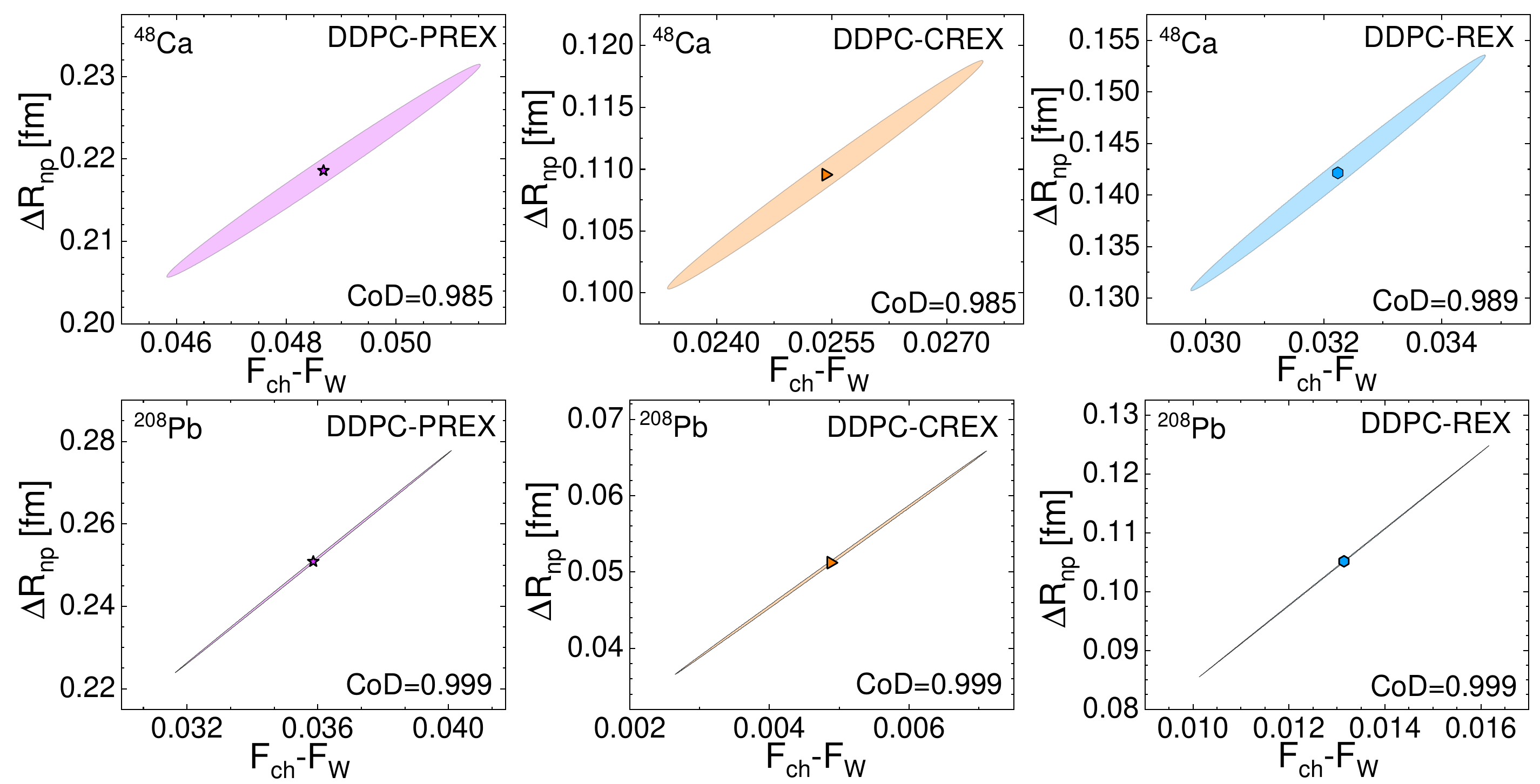}
	\caption{The error ellipsoids between the neutron skin thickness $\Delta R_{np}$ and the form factor difference $F_{ch}-F_W$ for $^{48}$Ca (upper panels) and $^{208}$Pb  (lower panels) using the new point-coupling interactions: DDPC-PREX, DDPC-CREX, and DDPC-REX. The CoD numbers are also given within the figures.}\label{fig:11}
\end{figure*}

The symmetry energy and its slope are known to be strongly correlated with the neutron skin thickness of nuclei \cite{PhysRevLett.85.5296,PhysRevC.86.015803,Tamii_2014,PhysRevC.81.051303}. However, accurate and model independent
measurement of the neutron skin thickness of nuclei is rather challenging.
The first Lead Radius EXperiment (PREX-I) \cite{PhysRevLett.108.112502}  estimated very large neutron skin thickness for $^{208}$Pb with significant 
uncertainties ($R_n-R_p = 0.33^{+0.16}_{-0.18}$). 
Recently, the new data from the PREX-II experiment has also been announced and the neutron-skin thickness of $^{208}$Pb was found as $R_{n}-R_{p}=$0.283$\pm$0.071 fm, obtained by combining the new data with the previous measurements \cite{PhysRevLett.126.172502}. The charge minus the weak form factor was also obtained as $F_{ch}-F_W=0.041\pm0.013$ in the same experiment.
Lately, CREX data has been announced and the neutron skin thickness and the form factor difference
 of $^{48}$Ca have been obtained as $R_{n}-R_{p}=$0.121$\pm$0.026(exp)$\pm$0.024(model) fm and $F_{ch}-F_W=0.0277\pm0.0055$, respectively \cite{PhysRevLett.129.042501}. In the following, we discuss the isovector properties of nuclei calculated using our three point coupling interactions constrained by $F_W$, in comparison to those of previously established EDFs and experimental data.

Figure \ref{fig:1} shows the neutron skin thickness $\Delta R_{np}$ for $^{48}$Ca and $^{208}$Pb as a function of the difference of charge and weak-charge form factors, $F_{ch}-F_W$, from calculations based on relativistic EDFs. The neutron skin thickness of a nucleus is calculated as
$\Delta R_{np}=\left\langle r_{n}^{2}\right\rangle^{1/2}-\langle r_{p}^{2}\rangle^{1/2}$, and $\langle r_{n(p)}^{2}\rangle^{1/2}$  represents neutron (proton) root-mean-square radii. The results of new point coupling functionals DDPC-CREX, DDPC-PREX, and DDPC-REX are compared with those of the DD-PC1 \cite{PhysRevC.78.034318}, DD-PCX \cite{PhysRevC.99.034318}, and family of eight DD-PC functionals that cover a range of the symmetry energy values at saturation density $J=29,30,...,36$ MeV \cite{Yuksel2021}, density dependent meson-exchange functionals 
DD-ME1 \cite{ddme1} and DD-ME2 \cite{ddme2}, and
the corresponding family of five DD-ME functionals with $J=30,32,...,38$ MeV.
The neutron skin thickness is strongly correlated with the $F_{ch}-F_W$ values, both for $^{48}$Ca 
and $^{208}$Pb. For $^{48}$Ca, only a few interactions are within the ranges of $F_{ch}-F_W$ and $\Delta R_{np}$ values obtained from the CREX experiment. As expected, the DDPC-CREX and DDPC-REX interactions
fit into expected range of the $F_{ch}-F_W$ and $\Delta R_{np}$ values,
since experimental value for $F_W(^{48}\text{Ca})$ 
is used in constraining their parameters.
However, all other interactions that have previously
been established as very successful in describing
nuclear properties, e.g. DD-ME1, DD-ME2, DD-PC1, DD-PCX, remain 
above the experimental ranges for $F_{ch}-F_W$ and $\Delta R_{np}$ values. The only exception is
DD-PC interaction with the $J=29$ value. In this work, the neutron skin thickness values for $^{48}$Ca are obtained as 0.218(5), 0.110(4), and 0.142(5) fm  using the DDPC-PREX, DDPC-CREX, and DDPC-REX interactions, respectively. We note that in the present optimization of the EDFs, 
the weak-charge form factor data with small adopted errors are used in order to
constrain the functionals that closely reproduce the experimental data on $F_W$, 
as previously introduced. Therefore, the symmetry energy and neutron-skin thickness values are both tightly constrained since these properties are strongly correlated with each other as well as with the weak-charge form factor data. Therefore, the obtained uncertainties in the neutron skin thickness values are small. It is seen that only the DDPC-REX result is in good agreement with the predictions of the \textit{ab initio} theory ($0.12 \leq \Delta R^{^{48}Ca}_{np} \leq 0.15$ fm) \cite{Hagen2016}, and the recent results by including triples corrections in coupled-cluster theory ($0.13 \leq \Delta R^{^{48}Ca}_{np} \leq 0.16$ fm)  \cite{Simonis2019}. On the other hand, it is lower compared to the \textit{model-averaged} values ($\Delta R^{^{48}Ca}_{np}=0.176 \pm  0.018$ fm) obtained in the EDF study in Ref. \cite{PhysRevC.85.041302}. Using the linear fit obtained with different EDFs and the experimental limits for $F_{ch}-F_W$ values, the neutron skin thickness is obtained between $0.098 \leq \Delta R^{^{48}Ca}_{np} \leq 0.147$ fm,
that is in better agreement with the \textit{ab initio} theory predictions \cite{Hagen2016,Simonis2019} than with the model-averaged EDF result \cite{PhysRevC.85.041302}.

 As shown in Fig. \ref{fig:1}, for $^{208}$Pb, the experimental errors in $F_{ch}-F_W$ and $\Delta R_{np}$ obtained from PREX-II are larger, and more interactions fit into the experimental range of their values. However, all these interactions are on the opposite side of those preferred from the CREX experiment, 
in agreement with their larger values of the symmetry energy parameter $J$. Among new functionals, only DDPC-PREX fits into the experimental range of $F_{ch}-F_W$ and $\Delta R_{np}$ values, while DDPC-CREX and DDPC-REX results are obtained at considerably lower values. The DD-PCX and DD-ME2 functionals also remain below the experimental range, while DD-ME1 and DD-PC1 ones are very close to the lower experimental limits
of $F_{ch}-F_W$ and $\Delta R_{np}$. From the families of interactions
that systematically vary the symmetry energy,
only those with $J \geq 34$ (DD-ME) and $J \geq 34$ (DD-PC) are consistent with the PREX-II data. Employing DDPC-PREX, DDPC-CREX, and DDPC-REX functionals in the calculations, the neutron skin thickness values for $^{208}$Pb are obtained as 0.250(10), 0.051(6), and 0.105(8) fm, respectively. 
The PREX-II measurements indicate a rather large neutron skin thickness value which is not in agreement neither with the \textit{model-averaged} values ($\Delta R^{^{208}Pb}_{np}=0.168 \pm  0.0022$ fm) obtained in Ref. \cite{PhysRevC.85.041302}, nor with the limits obtained using the \textit{ab initio} theory ($\Delta R^{^{208}Pb}_{np}=0.14-0.20$ fm) in Ref. \cite{https://doi.org/10.48550/arxiv.2112.01125}. Using the linear fit in Fig. \ref{fig:1} and experimental limits for $F_{ch}-F_W$ values, the neutron skin thickness is obtained between $0.202 \leq \Delta R^{^{208}Pb}_{np} \leq 0.371$ fm for $^{208}$Pb. Clearly, the limits are not in agreement with the previous model predictions \cite{PhysRevC.85.041302, https://doi.org/10.48550/arxiv.2112.01125}.

The uncertainties in the model parameters and observables, and the correlations between them can be obtained with the statistical covariance analysis \cite{Dobaczewski_2014}. Within this framework, the coefficient of determination (CoD) is used to determine whether two observables have a genuine statistical correlation (CoD) or not. The CoD values range between 0 and 1. Two quantities are strongly correlated when CoD = 1, as opposed to being uncorrelated when CoD = 0. Additionally, the error ellipsoid between the two observables can be shown as the graphical representation of the CoD (see Refs. \cite{Dobaczewski_2014,Erler_2015,Reinhard_2015}). 
In Figure \ref{fig:11}, we display the error ellipsoids between $\Delta R_{np}$ and $F_{ch}-F_W$ for $^{48}$Ca (upper panels) and $^{208}$Pb  (lower panels) using the new point-coupling functionals. The CoD numbers are also provided within the figures for each interaction and nucleus. Using three new point-coupling interactions, a strong correlation is obtained between the $\Delta R_{np}$ and $F_{ch}-F_W$ for both $^{48}$Ca and $^{208}$Pb. While the variances in the ellipsoids are slightly larger for $^{48}$Ca in the perpendicular direction, the error ellipsoids are quite narrow for $^{208}$Pb.
Similar results are also obtained in Ref. \cite{https://doi.org/10.48550/arxiv.2206.03134} using the non-relativistic functionals.

\begin{figure}
	\centering
	\includegraphics[width=\linewidth]{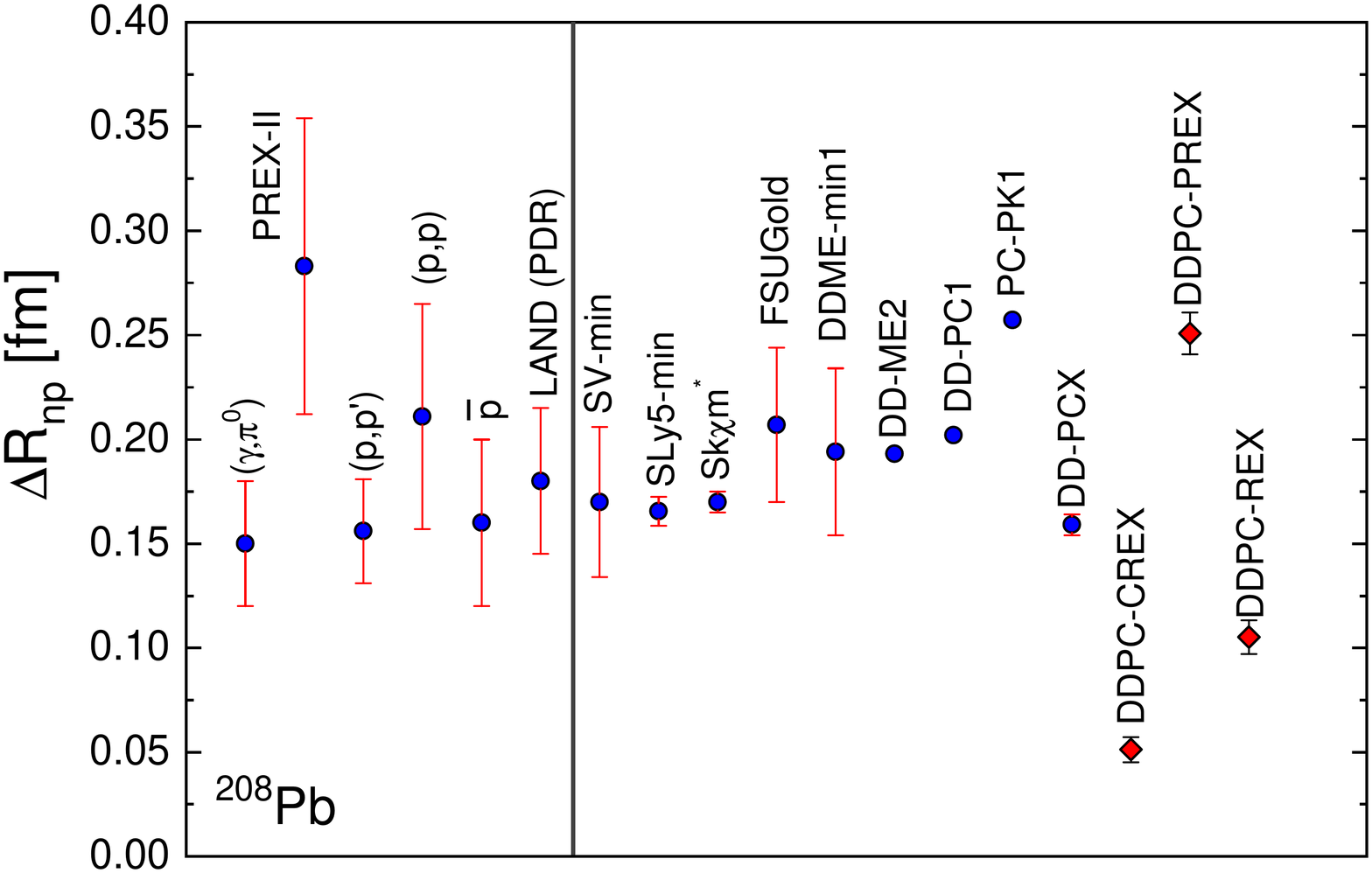}
	\caption{The neutron skin thickness of $^{208}$Pb for the DDPC-CREX, DDPC-PREX, and DDPC-REX functionals. For comparison, the experimental values are shown from ($\gamma$,$\pi^{0}$) \cite{tar14}, PREX-II \cite{PhysRevLett.126.172502}, ($p,p'$) \cite{PhysRevLett.107.062502}, ($p,p$) \cite{pp11}, $\overline{p}$ \cite{antip07}, LAND(PDR) \cite{klim07}. The results from the non-relativistic interactions include SV-min \cite{klup09}, SLy5-min~\cite{rocc15}, Sk$\chi$\textit{m}* \cite{Zhang2018}, and relativistic interactions: FSUGold \cite{Todd2005}, DDME-min1 \cite{rocc15}, DD-ME2 \cite{ddme2}, DD-PC1 \cite{PhysRevC.78.034318}, PC-PK1 \cite{PhysRevC.82.054319}, DD-PCX \cite{PhysRevC.99.034318}.
	}\label{fignp}
\end{figure}

Figure \ref{fignp} shows the summary of the neutron skin thicknesses values for $^{208}$Pb obtained using DDPC-CREX, DDPC-PREX, and DDPC-REX functionals,
in comparison to a number of previous experimental results \cite{tar14,PhysRevLett.126.172502,PhysRevLett.107.062502,pp11,antip07,klim07}. In addition, the results
of non-relativistic \cite{klup09,rocc15,Zhang2018} and relativistic \cite{Todd2005,rocc15,ddme2,PhysRevC.78.034318,PhysRevC.82.054319,PhysRevC.99.034318} studies based on the EDFs are shown, as well as the EDF constrained by chiral effective field theory, denoted
as Sk$\chi$\textit{m}*  \cite{Zhang2018} (more details on other EDFs are given in the figure caption). One can observe that for $^{208}$Pb the DDPC-CREX and DDPC-REX values for $\Delta R_{np}$ are smaller than those of all
previous measurements and EDF predictions. On the other side, the DDPC-PREX
functional provides $\Delta R_{np}$ value larger than obtained in
most of the previous studies. It is in agreement with 
the PREX-II experiment, but also with the PC-PK1 interaction which has a higher
value of the symmetry energy at saturation, $J=\text{35.6}$ MeV. 
Clearly, the data from the parity-violating electron scattering experiments CREX and PREX-II could not provide a consistent understanding of the neutron skin thickness in $^{208}$Pb. Recent study of 
parity violating asymmetry $A_{PV}$ in the EDF framework showed
that simultaneous accurate description of $A_{PV}$ in
$^{48}$Ca and $^{208}$Pb could also not be achieved \cite{https://doi.org/10.48550/arxiv.2206.03134}.

\begin{figure} [ht!]
	\centering
	\includegraphics[width=\linewidth]{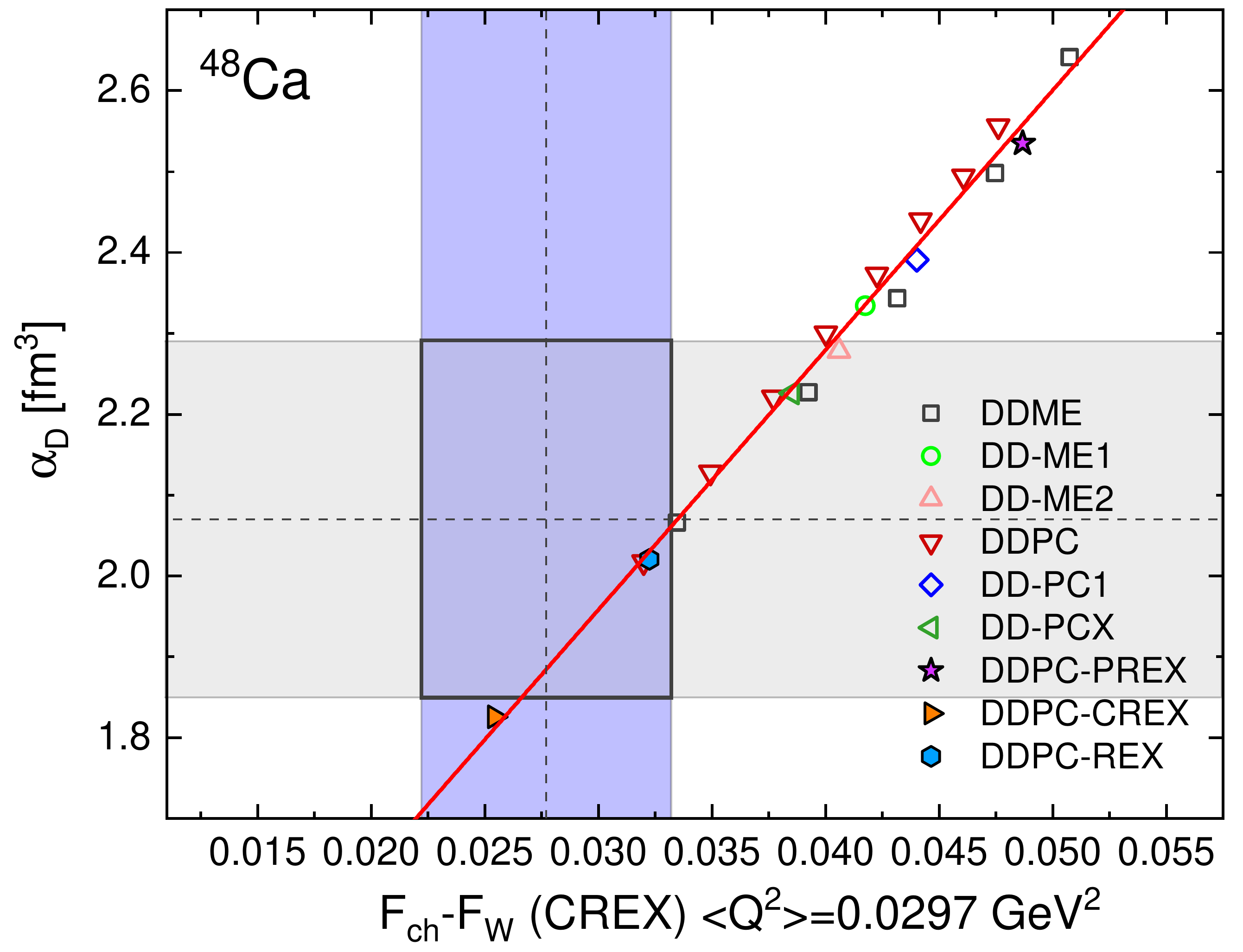}
	\includegraphics[width=\linewidth]{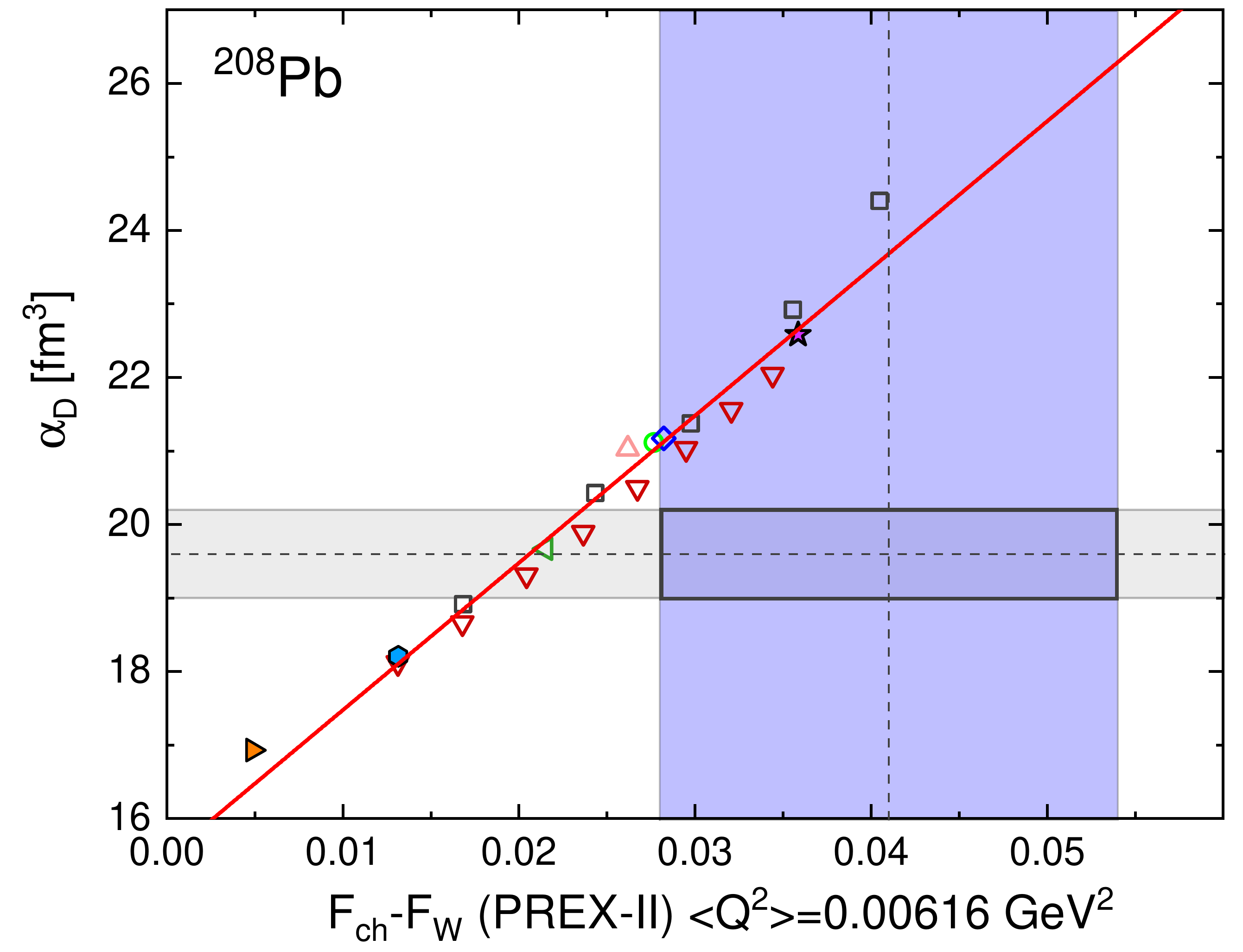}
	\caption{The dipole polarizability  $\alpha_D$ of $^{48}$Ca (upper panel) and $^{208}$Pb (lower panel) as a function of the form factor difference $F_{ch}-F_W$ using relativistic EDFs. Vertical bands denote $F_{ch}-F_W$ range of values from the CREX \cite{PhysRevLett.129.042501} and PREX-II \cite{PhysRevLett.126.172502} experiments, while horizontal bands correspond to the experimental data on $\alpha_D$ \cite{PhysRevLett.118.252501,PhysRevLett.107.062502,PhysRevC.92.064304}}\label{fig:2}
\end{figure}

\begin{figure*} [ht!]
	\centering
	\includegraphics[width=0.8\linewidth]{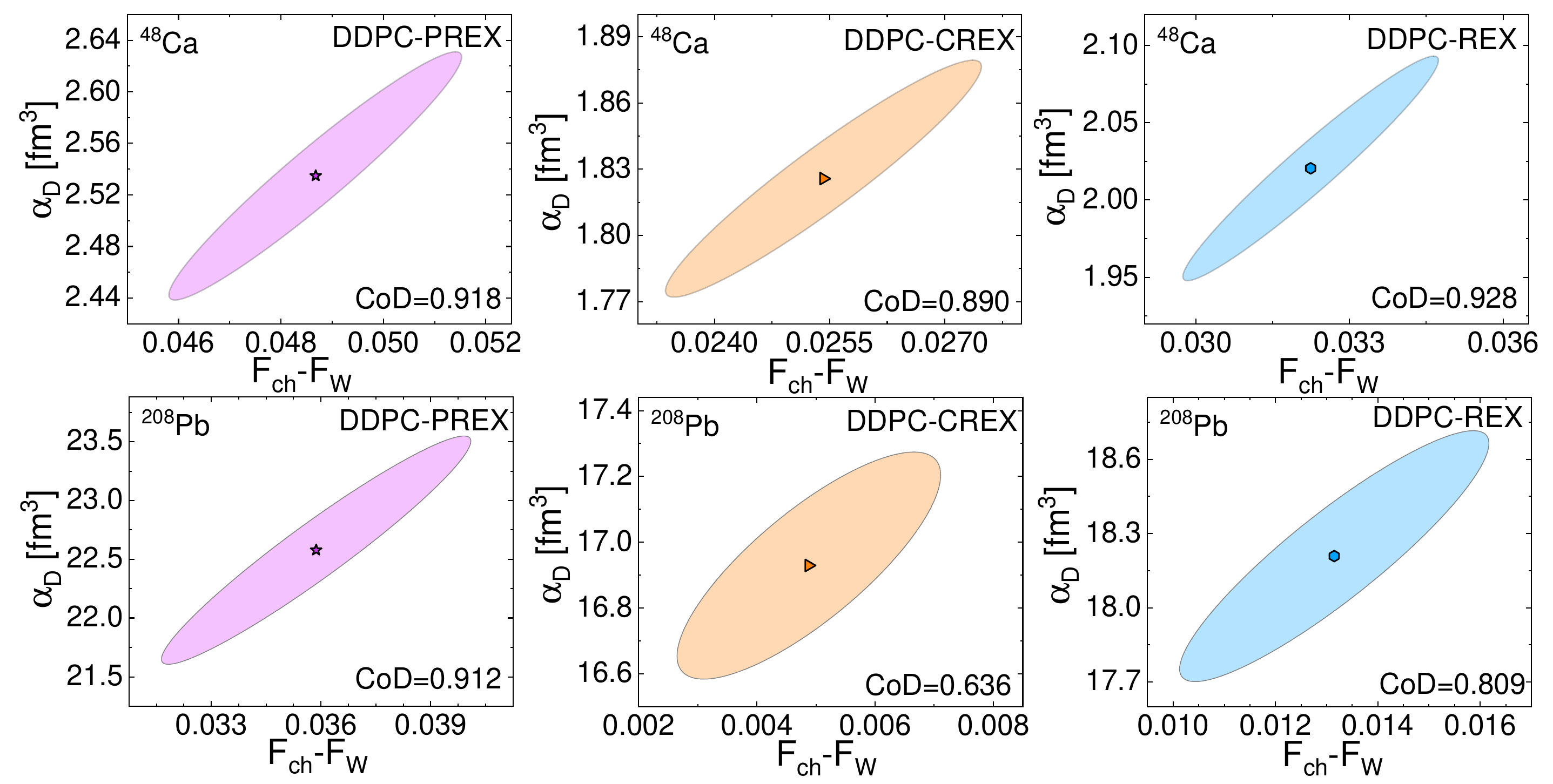}
	\caption{The error ellipsoids between the dipole polarizability $\alpha_D$ and form factor difference $F_{ch}-F_W$ using new point coupling interactions for  $^{48}$Ca (upper panels) and $^{208}$Pb (lower panels). The CoD values are also given within the figures.}\label{fig:12}
\end{figure*}

Another relevant quantity used in constraining the isovector channel 
of the EDFs is dipole polarizability $\alpha_D$, corresponding to the
sum of inverse energy weighted dipole transition strength in nuclei 
\cite{PhysRevC.81.051303}. Over the past decade the dipole polarizability
has attracted considerable interest because it is strongly correlated with 
the neutron form factor, neutron skin thickness  \cite{PhysRevC.81.051303,PhysRevC.85.041302,PhysRevC.88.024316,PhysRevC.92.064304}, and the
properties of the symmetry energy of nuclear matter \cite{PhysRevC.81.051303}.
Recently, the dipole polarizability has been measured for $^{48}$Ca \cite{PhysRevLett.118.252501} and $^{208}$Pb \cite{PhysRevLett.107.062502}, and the corresponding values to be used in analyses are 
$\alpha_D (^{48}\text{Ca})=$ 2.07$\pm$0.22 fm$^3$ \cite{PhysRevLett.118.252501} and $\alpha_D (^{208}\text{Pb})=$ 19.6$\pm$0.6 fm$^3$ \cite{PhysRevC.92.064304}. Therefore, it is interesting to confront the calculated values of $\alpha_D$ with those of $F_{ch}-F_W$,
together with the corresponding experimental data,
to assess the information which functional consistently describes these 
both quantities. The respective results are shown in Fig. \ref{fig:2},
using the same density-dependent point coupling and meson exchange functionals as previously discussed. The horizontal and vertical bands denote the experimental values with errors for $\alpha_D$ and $F_{ch}-F_W$, respectively. As expected, strong correlation between $\alpha_D$ and $F_{ch}-F_W$ is obtained for $^{48}$Ca and $^{208}$Pb for all
functionals used in the analysis. For $^{48}$Ca, only one
new interaction, DDPC-REX, and DDPC ($J$=29 MeV) are simultaneously within the 
experimental limits for $\alpha_D$ and $F_{ch}-F_W$ (CREX). 
The DDPC-CREX interaction gives a slightly smaller $\alpha_D$ value than the lower experimental limit, while the DDPC-PREX interaction
results for $\alpha_D$ and $F_{ch}-F_W$ are obtained at considerably
higher values and above the experimental range. 
The DD-PCX interaction,
which has been adjusted to reproduce experimental dipole
polarizabilities, seems not consistent with $F_{ch}-F_W$ (CREX). 
For $^{48}$Ca, the coupled cluster theory calculations predict $1.92 \leq \alpha_D (^{48}Ca) \leq 2.38$ fm$^{3}$, which is in agreement only with predictions of the DDPC-REX interaction \cite{Simonis2019}.

Lower panel of Fig. \ref{fig:2}, shows that for 
$^{208}$Pb none of the new or previously established functionals 
can simultaneously reproduce experimental 
values for $\alpha_D$ and $F_{ch}-F_W$ (PREX-II). As expected, DDPC-PREX reproduces experimental form factors,
but considerably overestimates $\alpha_D$ values. 
The DDPC-CREX and DDPC-REX underestimate both
$\alpha_D$ and $F_{ch}-F_W$ (PREX-II) values. There
is a considerable lack of consistency between the functionals
constrained using CREX and PREX-II data when confronted with
dipole polarizability studies. In Ref. \cite{https://doi.org/10.48550/arxiv.2206.03134}, similar calculations have been performed using the non-relativistic functionals, and a strong correlation is also  obtained between the parity violating asymmetry $A_{PV}$ and $J\alpha_D$. Since $A_{PV}$ is directly correlated with the $F_{ch}$ and $F_W$, our findings are consistent.

Figure \ref{fig:12} shows the error ellipsoids between the $\alpha_D$ and $F_{ch}-F_W$ for the new point-coupling interactions. The results are displayed for $^{48}$Ca (upper panels) and $^{208}$Pb (lower panels). For both nuclei, there is a strong correlation between $\alpha_D$ and $F_{ch}-F_W$ using three functionals, albeit smaller when compared to $\Delta R_{np}$ and $F_{ch}-F_W$. The lowest CoD value between the $\alpha_D$ and $F_{ch}-F_W$ is obtained for $^{208}$Pb using the DDPC-CREX functional.

In summary, the weak-charge form factors from parity-violating electron
scattering open an important new perspective to constrain the neutron skins and 
the properties of the nuclear matter symmetry energy.
In this work, the weak-charge form factors obtained from the CREX and PREX-II data have been used directly in constraining the relativistic EDFs, having in this way a considerable impact on 
their isovector interaction channel. However, the neutron skin
thicknesses calculated from these EDFs do not seem consistent, and they are
at variance with respect to previous theoretical and experimental studies.
When the weak-charge form factors are confronted with the dipole polarizability,
the results of the analysis show significant inconsistencies, although both
should probe the same content of the EDFs. The symmetry energy and its slope
at saturation density, as well as the neutron skin thickness, are significantly smaller for the DDPC-CREX functional in comparison to those of the
DDPC-PREX, and values established by previous EDFs are between those obtained with the two new functionals.
The analysis of this work
shows that no consistent conclusions from the theoretical side
could be obtained when using recent CREX and PREX-II results. Clearly, further
EDF and \textit{ab-initio} studies, alongside with novel experimental investigations are
needed to resolve the current puzzling implications of the parity violating electron
scattering data.

\section*{Acknowledgements}
This work is supported by the QuantiXLie Centre of Excellence, a project co financed by the Croatian Government and European Union through the European Regional Development Fund, the Competitiveness and Cohesion Operational Programme (KK.01.1.1.01.0004). 
\bibliographystyle{apsrev4-2}
\bibliography{DDPCREX}
\end{document}